\documentclass{article}

\usepackage{microtype}
\usepackage{graphicx}
\usepackage{subfigure}
\usepackage{booktabs} 

\usepackage{hyperref}

\usepackage[accepted]{icml2025}

\usepackage{amsmath}
\usepackage{amssymb}
\usepackage{mathtools}
\usepackage{amsthm}

\usepackage[capitalize,noabbrev]{cleveref}

\theoremstyle{plain}
\newtheorem{theorem}{Theorem}[section]
\newtheorem{proposition}[theorem]{Proposition}

\theoremstyle{definition}

\theoremstyle{remark}

\usepackage[textsize=tiny]{todonotes}

\icmltitlerunning{Submission and Formatting Instructions for ICML 2025}

\begin{document}

\twocolumn[
\icmltitle{Nemesis: Noise-randomized Encryption with Modular Efficiency and Secure Integration in Machine Learning Systems}



\icmlsetsymbol{equal}{*}

\begin{icmlauthorlist}
\icmlauthor{Dongfang Zhao}{uw}
\end{icmlauthorlist}

\icmlaffiliation{uw}{University of Washington, United States}


\icmlcorrespondingauthor{Dongfang Zhao}{dzhao@cs.washington.edu}

\icmlkeywords{Federated Learning, Homomorphic Encryption, Caching}

\vskip 0.3in
]



\printAffiliationsAndNotice{}  

\begin{abstract}
Machine learning (ML) systems that guarantee security and privacy often rely on Fully Homomorphic Encryption (FHE) as a cornerstone technique, enabling computations on encrypted data without exposing sensitive information. However, a critical limitation of FHE is its computational inefficiency, making it impractical for large-scale applications. In this work, we propose \textit{Nemesis}, a framework that accelerates FHE-based systems without compromising accuracy or security. The design of Nemesis is inspired by Rache (SIGMOD'23), which introduced a caching mechanism for encrypted integers and scalars. Nemesis extends this idea with more advanced caching techniques and mathematical tools, enabling efficient operations over multi-slot FHE schemes and overcoming Rache's limitations to support general plaintext structures. We formally prove the security of Nemesis under standard cryptographic assumptions and evaluate its performance extensively on widely used datasets, including MNIST, FashionMNIST, and CIFAR-10. Experimental results show that Nemesis significantly reduces the computational overhead of FHE-based ML systems, paving the way for broader adoption of privacy-preserving technologies.
\end{abstract}

\section{Introduction}

Privacy-preserving machine learning (PPML)~\cite{pjain_icml13} has become a critical research area, driven by the need to protect sensitive data in applications ranging from healthcare to finance. Fully Homomorphic Encryption (FHE)~\cite{gentry2009fhe} is a prominent approach in this domain, allowing computations to be performed directly on encrypted data without decrypting it, thus maintaining data confidentiality. FHE is particularly advantageous compared to other techniques like Multi-Party Computation (MPC)~\cite{mpc} and Differential Privacy (DP)~\cite{dwork2006dp}. While MPC requires multiple parties to participate in interactive protocols, FHE enables standalone computation on encrypted data. Similarly, unlike DP, which injects noise into the data or results, FHE preserves the exactness of computations, making it an attractive choice for high-accuracy ML systems. This has led to widespread adoption of FHE in privacy-preserving systems, including federated learning and secure inference.
Popular FHE libraries include OpenFHE~\cite{OpenFHE}, HELib~\cite{helib}, Microsoft SEAL~\cite{seal}, TenSEAL~\cite{tenseal}, and Lattigo~\cite{lattigo}.

Despite its promise, FHE suffers from a significant drawback: it is computationally expensive, often orders of magnitude slower than unencrypted operations~\cite{fhe_acc_survey}. This inefficiency has spurred research into optimizing FHE for machine learning systems, particularly for operations like encryption and ciphertext manipulation~\cite{eaha_sp22}. Recent works have proposed various techniques, such as batch processing and algorithmic improvements, to mitigate these costs. Among them, Rache~\cite{otawose_sigmod23} introduced a caching mechanism to optimize scalar encryption in FHE, providing substantial performance gains. However, Rache's design is limited to integer or scalar plaintexts, making it unsuitable for more general use cases like multi-slot FHE schemes, which are widely used in ML systems for encrypting and processing multiple plaintexts in parallel.

Motivated by these challenges, we introduce \textit{Nemesis}, a framework designed to extend caching-based optimizations to multi-slot FHE. Nemesis aims to support efficient precomputation for batch encryption, ensuring high performance while maintaining the semantic security of the underlying FHE scheme. By leveraging advanced mathematical tools and a novel caching strategy, Nemesis bridges the gap left by previous works, making it a practical solution for real-world PPML applications. This work not only addresses a key limitation of FHE but also establishes a foundation for broader adoption of privacy-preserving technologies in machine learning.

The contributions of this paper are summarized as follows.
\begin{itemize}
    \item First caching framework for multi-slot FHE: We propose \textit{Nemesis}, the first optimization framework designed to cache and reuse computations in multi-slot FHE systems. By reducing redundant operations, Nemesis substantially improves the efficiency of privacy-preserving machine learning systems, such as those leveraging federated learning.
    
    \item Theoretical security guarantees: We provide a formal security analysis of Nemesis, proving its semantic security under the IND-CPA assumptions of the underlying FHE scheme. This includes a detailed reduction-based proof, ensuring the correctness and robustness of the method.
    
    \item Comprehensive experimental evaluation: We conduct extensive experiments using publicly available implementations, including the federated learning framework \texttt{PFLlib} and the state-of-the-art homomorphic encryption library \texttt{OpenFHE}. The evaluations are performed on widely recognized datasets such as MNIST, FashionMNIST, and CIFAR-10, demonstrating the practical benefits of Nemesis.
\end{itemize}


\section{Related Work and Preliminaries}

Privacy-preserving machine learning (PPML)~\cite{pjain_icml13} has gained significant attention as a means to ensure data confidentiality while enabling collaborative training or inference on sensitive datasets. Techniques such as Multi-Party Computation (MPC)~\cite{mpc}, Differential Privacy (DP)~\cite{dwork2006dp}, and Fully Homomorphic Encryption (FHE)~\cite{gentry2009fhe} have been extensively explored in the literature. MPC allows multiple parties to jointly compute functions over their inputs without revealing the inputs themselves, but it often requires high communication overhead and interactive protocols. DP, on the other hand, provides privacy guarantees by injecting noise into data or computation results, potentially compromising accuracy. FHE stands out in this landscape as it enables exact computation on encrypted data without requiring interaction or introducing noise, making it particularly suitable for high-accuracy ML tasks. Recent advancements have integrated FHE into PPML systems for applications like secure federated learning and private inference. However, these systems still face significant performance challenges, particularly in encryption and ciphertext operations, where further optimizations are needed to make PPML scalable and practical.

FHE enables computations to be performed directly on encrypted data while preserving the confidentiality of the underlying plaintexts. Unlike traditional encryption schemes, FHE supports operations like addition and multiplication on ciphertexts, which correspond to the same operations on plaintexts. This property, known as homomorphism, makes FHE a powerful tool for secure computations in privacy-sensitive domains. Modern FHE schemes, such as BFV~\cite{bfv}, CKKS~\cite{ckks}, and TFHE~\cite{chillotti2016tfhe}, employ various optimizations to improve computational efficiency and support different types of data, from integers to approximate floating-point values. However, even with these advancements, FHE remains computationally expensive, with encryption often being the most resource-intensive step. To address this, recent works have proposed techniques such as batching, approximate arithmetic, and precomputation. Among these, Rache~\cite{otawose_sigmod23} introduced a caching mechanism for scalar encryption, reducing redundancy but limited to simple plaintext structures. These limitations motivate the need for more generalized solutions like Nemesis, which extend caching optimizations to multi-slot FHE schemes and broader plaintext use cases.

A fundamental aspect of cryptographic schemes is their formal security analysis. The most widely accepted standard for security is indistinguishability under chosen plaintext attack (IND-CPA)~\cite{goldwasser1982indcpa}, which ensures that an adversary cannot distinguish between encryptions of two chosen plaintexts. Security proofs typically involve demonstrating that breaking the scheme would imply solving an underlying hard problem, such as factoring large integers or computing discrete logarithms. This is achieved through a reduction, where the adversary's success is transformed into an efficient algorithm for solving the hard problem. In the context of FHE, semantic security guarantees that the ciphertext reveals no information about the plaintext, even under adaptive attacks. 

\section{Methodology}

The Nemesis framework is designed to enhance the efficiency of encryption in fully homomorphic encryption (FHE) systems, particularly in the context of machine learning. The approach builds on three key components: precomputation, reconstruction, and randomization. Precomputation minimizes redundant computations by generating reusable encrypted scalar values, which can be applied across different encryption tasks. Reconstruction involves efficient encoding of plaintexts into polynomials, ensuring that multiple plaintexts can be compactly represented and manipulated while maintaining correctness. Randomization introduces controlled noise directly into ciphertexts, balancing security and efficiency by ensuring that the ciphertext remains indistinguishable without introducing significant computational overhead. These components collectively optimize the overall encryption process. Nemesis is implemented on top of the CKKS scheme, utilizing its approximate arithmetic capabilities for scalability and precision in encrypted computations. By integrating these techniques, Nemesis achieves efficient encryption for real-world applications, particularly those requiring high-throughput computation.

\subsection{Precomputation in Nemesis}

Nemesis' precomputation is described in Algorithm~\ref{alg:precompute}. The precomputation stage in Nemesis optimizes encryption by encoding a carefully chosen batch of plaintexts into a polynomial using roots of unity and leveraging a Vandermonde matrix structure. The goal is to generate a precomputed ciphertext that can be reused across multiple computations, reducing runtime overhead.

\begin{algorithm}[h]
\caption{Precomputing a Polynomial with Vandermonde Encoding}
\label{alg:precompute}
\begin{algorithmic}[1]
\INPUT Plaintext candidates \(\mathcal{M}_{\text{candidates}} = \{m_0, m_1, \ldots\}\), encryption function \(\text{Enc}(\cdot)\), public key \(\text{pk}\), target polynomial degree \(d\)
\OUTPUT Precomputed ciphertext \(\text{Ciphertext}\)
\STATE Select \(d\) plaintexts \(\{m_0, m_1, \ldots, m_{d-1}\} \subset \mathcal{M}_{\text{candidates}}\) based on frequency of use or application needs
\STATE Compute a primitive \(d\)-th root of unity \(\omega\) in the plaintext space
\STATE Construct the \(d \times d\) Vandermonde matrix \(\mathbf{V}\) with \(V_{ij} = (\omega^j)^i\) for \(i, j = 0, \ldots, d-1\)
\STATE Solve the linear system \(\mathbf{V} \cdot \mathbf{m} = \mathbf{c}\) to compute the polynomial coefficients \(\mathbf{c} = [c_0, c_1, \ldots, c_{d-1}]^\top\)
\STATE Encode the polynomial \(\mathcal{P}(x) \gets \sum_{i=0}^{d-1} c_i \cdot x^i\)
\STATE Encrypt the encoded polynomial:
\[
\text{Ciphertext} \gets \text{Enc}_{\text{pk}}(\mathcal{P}(x))
\]
\STATE Return the precomputed ciphertext \(\text{Ciphertext}\)
\end{algorithmic}
\end{algorithm}

Given a batch of plaintexts \(\{m_0, m_1, \ldots, m_{d-1}\}\), the encoding process first selects these plaintexts based on their frequency of use or computational relevance. Once selected, the plaintexts are mapped into a polynomial \(\mathcal{P}(x)\) over the plaintext space using a Vandermonde structure:
\[
\mathcal{P}(x) = \sum_{i=0}^{d-1} m_i \cdot x^i,
\]
where \(x\) is the formal polynomial variable (i.e., \textit{indeterminate}). This process can be represented as solving a linear system:
\[
\mathbf{V} \cdot \mathbf{m} = \mathbf{c},
\]
where \(\mathbf{V}\) is a \(d \times d\) Vandermonde matrix with entries \(V_{ij} = (\omega^j)^i\), \(\mathbf{m} = [m_0, m_1, \ldots, m_{d-1}]^\top\) is the vector of plaintexts, and \(\mathbf{c} = [c_0, c_1, \ldots, c_{d-1}]^\top\) represents the polynomial coefficients.

The encoded polynomial \(\mathcal{P}(x)\) is then encrypted using the encryption function \(\text{Enc}(\cdot)\) and the public key pk:
\[
\text{Ciphertext} \gets \text{Enc}_{\text{pk}}(\mathcal{P}(x)).
\]

The choice of \(\{m_0, m_1, \ldots, m_{d-1}\}\) ensures that the encoded polynomial efficiently supports typical homomorphic operations, making this precomputed ciphertext a reusable and efficient building block.

\subsection{Reconstruction in Nemesis}

The reconstruction stage in Nemesis is designed to leverage the precomputed ciphertext to efficiently encrypt target plaintext batches without re-performing computationally expensive encryption from scratch. This approach is particularly advantageous in scenarios where repeated encryption of varying plaintexts with a fixed polynomial structure is required, such as secure machine learning or encrypted database queries.
Given a precomputed ciphertext \(\text{Ciphertext}_{\text{base}} = \text{Enc}(\mathcal{P}_{\text{base}}(x))\), where \(\mathcal{P}_{\text{base}}(x) = \sum_{i=0}^{d-1} b_i \cdot x^i\), and a target plaintext batch \(\{m_0, m_1, \ldots, m_{d-1}\}\), the reconstruction process is described in Algorithm~\ref{alg:reconstruct}.

\begin{algorithm}[h]
\caption{Reconstructing Ciphertext with Dynamic Scaling Factors}
\label{alg:reconstruct}
\begin{algorithmic}[1]
\INPUT Precomputed ciphertext \(\text{Ciphertext}_{\text{base}} = \text{Enc}(\mathcal{P}_{\text{base}}(x))\), target plaintext batch \(\{m_0, m_1, \ldots, m_{d-1}\}\)
\OUTPUT Reconstructed ciphertext \(\text{Ciphertext}_{\text{target}}\)
\STATE Extract coefficients of \(\mathcal{P}_{\text{base}}(x)\): \(\{b_0, b_1, \ldots, b_{d-1}\}\)
\STATE Initialize a list of scaled plaintexts: \( \text{scaled\_values} = [] \)
\FOR{\(i = 0\) to \(d-1\)}
    \STATE Compute the scaling factor:
    \[
    s_i \gets b_i
    \]
    \STATE Scale the plaintext value:
    \[
    m'_i \gets \frac{m_i}{s_i}
    \]
    \STATE Append \(m'_i\) to \( \text{scaled\_values} \)
\ENDFOR
\STATE Construct the polynomial \(\mathcal{P}'(x)\) from \( \text{scaled\_values} \):
\[
\mathcal{P}'(x) \gets \sum_{i=0}^{d-1} m'_i \cdot x^i
\]
\STATE Perform homomorphic multiplication with the precomputed ciphertext:
\[
\text{Ciphertext}_{\text{target}} \gets \text{Ciphertext}_{\text{base}} \circledast \mathcal{P}'(x)
\]
\STATE Return reconstructed ciphertext \(\text{Ciphertext}_{\text{target}}\)
\end{algorithmic}
\end{algorithm}

The process begins with the scaling of each plaintext value based on the coefficients of the precomputed ciphertext. These scaling factors are derived to ensure that the resulting encoded polynomial aligns with the precomputed structure. Specifically, given the precomputed ciphertext \(\text{Ciphertext}_{\text{base}} = \text{Enc}(\mathcal{P}_{\text{base}}(x))\), which encrypts a polynomial \(\mathcal{P}_{\text{base}}(x) = \sum_{i=0}^{d-1} b_i \cdot x^i\), each target plaintext \(m_i\) is scaled by its corresponding coefficient \(b_i\):
\[
m'_i = \frac{m_i}{b_i}.
\]
This step ensures that the scaled plaintext values \(m'_i\) are compatible with the structure of \(\mathcal{P}_{\text{base}}(x)\).

Once scaled, the plaintext batch is encoded into a polynomial \(\mathcal{P}'(x) = \sum_{i=0}^{d-1} m'_i \cdot x^i\). This encoding mirrors the approach used in the precomputation stage, utilizing the polynomial ring defined by the encryption scheme's parameters. The coefficients of \(\mathcal{P}'(x)\) are represented within the modulus space \(q\), ensuring compatibility with the homomorphic encryption scheme.

The final step involves performing homomorphic multiplication between the precomputed ciphertext and the encoded polynomial:
\[
\text{Ciphertext}_{\text{target}} = \text{Ciphertext}_{\text{base}} \circledast \mathcal{P}'(x),
\]
where \(\circledast\) denotes the special homomorphic multiplication operation. This operation inherently applies modular reduction to ensure the ciphertext coefficients remain within the valid range defined by \(q\), while preserving the additive and multiplicative homomorphic properties of the encryption scheme.

\subsection{Randomization in Nemesis}

The randomization stage in Nemesis introduces controlled noise into the reconstructed ciphertext to enhance security. By adding a small random polynomial to the output of the reconstruction process, this stage prevents deterministic patterns in the ciphertext, which could otherwise expose sensitive information to adversaries. The randomization step preserves the correctness of the decryption process while increasing the difficulty of statistical attacks. The full algorithm is described in Algorithm~\ref{alg:randomization}.

\begin{algorithm}[h]
\caption{Randomizing Reconstructed Ciphertexts}
\label{alg:randomization}
\begin{algorithmic}[1]
\INPUT Reconstructed ciphertext \(\text{Ciphertext}_{\text{target}} = \text{Enc}(\mathcal{P}'(x))\), randomization standard deviation \(\sigma\)
\OUTPUT Randomized ciphertext \(\text{Ciphertext}_{\text{randomized}}\)
\STATE Extract parameters of the polynomial ring, including modulus \(q\) and degree \(d\)
\STATE Initialize a random polynomial \(\mathcal{R}(x) \gets 0\)
\FOR{\(i = 0\) to \(d-1\)}
    \STATE Sample a random coefficient \(r_i \sim \mathcal{N}(0, \sigma^2)\)
    \STATE Round \(r_i\) to the nearest integer and reduce modulo \(q\):
    \[
    r_i \gets \text{round}(r_i) \mod q
    \]
    \STATE Update the polynomial:
    \[
    \mathcal{R}(x) \gets \mathcal{R}(x) + r_i \cdot x^i
    \]
\ENDFOR
\STATE Add the random polynomial to the ciphertext via homomorphic addition:
\[
\text{Ciphertext}_{\text{randomized}} \gets \text{Ciphertext}_{\text{target}} \oplus \mathcal{R}(x)
\]
\STATE Return randomized ciphertext \(\text{Ciphertext}_{\text{randomized}}\)
\end{algorithmic}
\end{algorithm}

Given the reconstructed ciphertext \(\text{Ciphertext}_{\text{target}} = \text{Enc}(\mathcal{P}'(x))\), where \(\mathcal{P}'(x)\) is the encoded polynomial of the target plaintext batch, the randomization process generates a random polynomial \(\mathcal{R}(x)\) with coefficients drawn from a discrete Gaussian distribution:
\[
\mathcal{R}(x) = \sum_{i=0}^{d-1} r_i \cdot x^i,
\]
where \(r_i \sim \mathcal{N}(0, \sigma^2)\) and \(\sigma\) is the standard deviation of the Gaussian distribution. The degree \(d\) of \(\mathcal{R}(x)\) matches that of \(\mathcal{P}'(x)\), and the coefficients are bounded by the plaintext modulus \(q\) to ensure compatibility with the encryption scheme.

The randomized ciphertext is computed by directly applying homomorphic addition between the reconstructed ciphertext and the random polynomial:
\[
\text{Ciphertext}_{\text{randomized}} = \text{Ciphertext}_{\text{target}} \oplus \mathcal{R}(x),
\]
where \(\oplus\) denotes homomorphic addition, ensuring both modular reduction and ciphertext alignment.

\subsection{Security Analysis}

\subsubsection{Threat Model}

In our security analysis, we assume a standard threat model in the context of fully homomorphic encryption (FHE). The adversary is computationally bounded and aims to compromise the confidentiality of encrypted data by leveraging access to ciphertexts, observing patterns in encrypted computations, or exploiting weaknesses in the randomization process. The following assumptions define the boundaries of our threat model:

\begin{itemize}
    \item The underlying FHE scheme is semantically secure under chosen plaintext attacks (IND-CPA).
    \item The adversary has no access to the secret key and cannot modify or inject arbitrary ciphertexts.
    \item Side-channel attacks, such as timing or power analysis, are beyond the scope of this work.
    \item The noise growth within the ciphertexts remains within the decryptable range throughout the computation.
\end{itemize}

Our analysis focuses on ensuring that even with repeated access to ciphertexts generated by the Nemesis scheme (e.g., from reconstruction and randomization), the adversary cannot infer meaningful information about the underlying plaintexts.

\subsubsection{Security Proof}

We now formally analyze the security of Nemesis under the above threat model. The proof demonstrates that Nemesis satisfies IND-CPA security by reducing its security to that of the underlying FHE scheme.

Let \(\text{Adv}_{\mathcal{A}}\) denote the advantage of an adversary \(\mathcal{A}\) in distinguishing encryptions under Nemesis. For any adversary \(\mathcal{A}\), \(\text{Adv}_{\mathcal{A}}\) is defined as:
\[
\begin{aligned}
\text{Adv}_{\mathcal{A}} = &\left| \Pr[\mathcal{A}(\text{Ciphertext}_{0}) = 1] \right. \\
&\left. - \Pr[\mathcal{A}(\text{Ciphertext}_{1}) = 1] \right|,
\end{aligned}
\]
where \(\text{Ciphertext}_0\) and \(\text{Ciphertext}_1\) are encryptions of two plaintexts \(m_0\) and \(m_1\) under Nemesis.
A scheme is IND-CPA secure if for any probabilistic polynomial-time (PPT) adversary \(\mathcal{A}\), \(\text{Adv}_{\mathcal{A}}\) is negligible, i.e., there exists a negligible function \(\epsilon(\lambda)\) such that \(\text{Adv}_{\mathcal{A}} \leq \epsilon(\lambda)\), where \(\lambda\) is the security parameter.

\begin{proposition}
Under the assumption that the underlying FHE scheme is IND-CPA secure, Nemesis maintains semantic security for all ciphertexts generated by the scheme.    
\end{proposition}

\begin{proof}
The security of Nemesis is established by demonstrating that each stage (precomputation, reconstruction, and randomization) preserves semantic security under the assumption that the underlying FHE scheme is IND-CPA secure. The overall security then follows from the composability of IND-CPA security in the FHE setting.

In the precomputation stage, Nemesis encodes a carefully chosen batch of plaintexts into a polynomial \(\mathcal{P}_{\text{base}}(x)\) and encrypts it to generate \(\text{Ciphertext}_{\text{base}} = \text{Enc}(\mathcal{P}_{\text{base}}(x))\). By the IND-CPA security of the FHE scheme, the ciphertext \(\text{Ciphertext}_{\text{base}}\) is computationally indistinguishable from a random ciphertext:
\[
\left| \Pr[\mathcal{A}(\text{Ciphertext}_{\text{base}}) = 1] - \Pr[\mathcal{A}(r) = 1] \right| \leq \epsilon_1(\lambda),
\]
where \(r\) is a randomly chosen ciphertext and \(\epsilon_1(\lambda)\) is negligible. Thus, observing \(\text{Ciphertext}_{\text{base}}\) provides no advantage to the adversary in distinguishing the underlying plaintext polynomial \(\mathcal{P}_{\text{base}}(x)\).

In the reconstruction stage, the target plaintexts \(\{m_0', m_1', \ldots, m_{d-1}'\}\) are scaled and encoded into a new polynomial \(\mathcal{P}'(x)\), which is homomorphically combined with \(\text{Ciphertext}_{\text{base}}\) to generate:
\[
\text{Ciphertext}_{\text{target}} = \text{Ciphertext}_{\text{base}} \circledast \mathcal{P}'(x),
\]
where \(\circledast\) represents the homomorphic multiplication operation with modular reduction. By the correctness of the homomorphic encryption scheme, this operation does not expose any information about \(\mathcal{P}'(x)\) beyond what is already protected by \(\text{Ciphertext}_{\text{base}}\).

Formally, there exists a reduction \(\mathcal{B}\) to the security of the FHE scheme such that for any adversary \(\mathcal{A}\) attempting to distinguish \(\text{Ciphertext}_{\text{target}}\):
\[
\text{Adv}_{\mathcal{A}} \leq \text{Adv}_{\mathcal{B}} + \epsilon_2(\lambda),
\]
where \(\epsilon_2(\lambda)\) is negligible. Therefore, the semantic security of the reconstructed ciphertext \(\text{Ciphertext}_{\text{target}}\) is preserved.

In terms of reduction efficiency, there exists a polynomial-time reduction \(\mathcal{B}\) that transforms any adversary \(\mathcal{A}\) attacking Nemesis into an adversary attacking the underlying FHE scheme. In each stage of Nemesis:

\begin{itemize}
    \item Precomputation: Encoding the plaintexts into a polynomial \(\mathcal{P}_{\text{base}}(x)\) involves \(O(d^2)\) operations due to matrix-vector multiplication. Encryption is a single invocation of the base FHE scheme.
    \item Reconstruction: Scaling, encoding, and homomorphic multiplication each have complexity bounded by \(O(d)\) for \(d\) plaintext slots.
    \item Randomization: Generating a random polynomial \(\mathcal{R}(x)\) and performing homomorphic addition both have complexity \(O(d)\).
\end{itemize}

Each operation in Nemesis is bounded by polynomial time in the security parameter \(\lambda\) and plaintext batch size \(d\). Hence, the reduction \(\mathcal{B}\) runs in polynomial time, ensuring that the semantic security of Nemesis is directly reducible to the IND-CPA security of the underlying FHE scheme.

By combining the above results, the overall advantage of an adversary \(\mathcal{A}\) in distinguishing encryptions under Nemesis is bounded by:
\[
\text{Adv}_{\mathcal{A}} \leq \epsilon_1(\lambda) + \epsilon_2(\lambda) + \epsilon_3(\lambda).
\]
Since each term is negligible, the total advantage \(\text{Adv}_{\mathcal{A}}\) is also negligible.
Therefore, under the assumption that the underlying FHE scheme is IND-CPA secure, Nemesis maintains semantic security for all ciphertexts generated by the scheme.
\end{proof}

\section{Evaluation}

\subsection{System Implementation}

Nemesis is implemented upon two existing frameworks, \textit{PFLlib}~\cite{PFLlib} and \textit{OpenFHE}~\cite{OpenFHE}, making significant enhancements to meet the requirements of our extended Nemesis system. Specifically, we have contributed over 3,000 lines of source code modifications, including more than 200 lines of Python code in \textit{PFLlib} and over 2,800 lines of C++ code in \textit{OpenFHE}. These modifications involve implementing new algorithms, optimizing data processing pipelines, enabling dynamic ciphertext operations, and improving polynomial noise randomization and batch processing. 

\paragraph{PFLlib}~\cite{PFLlib}
We use this library for machine learning models and datasets. \textit{PFLlib} is a comprehensive framework designed for privacy-preserving federated learning, providing ready-to-use models and standardized datasets such as MNIST, FashionMNIST, and CIFAR-10. In this project, we have made over 200 lines of source code modifications (mostly Python) to enhance the framework's functionality, including implementing new algorithms, optimizing data processing pipelines, and integrating additional performance benchmarks. These changes have been committed to a dedicated branch, ensuring compatibility with our extended Nemesis system and enabling seamless experimentation with dynamic batch sizes and time evaluations.

\paragraph{OpenFHE}~\cite{OpenFHE}
Our fully homomorphic encryption (FHE) implementation is based on the \textit{OpenFHE} library. \textit{OpenFHE} is a state-of-the-art open-source library for lattice-based cryptography, supporting various FHE schemes such as BFV, BGV, and CKKS. It provides a modular and efficient implementation, which we extend to include custom optimizations for our encryption schemes. In this project, we have made over 2,800 lines of source code modifications, primarily in C++, to integrate dynamic ciphertext operations, optimize polynomial noise randomization, and enable efficient batch processing. These enhancements are implemented in a new branch to ensure modularity, maintainability, and compatibility with the original library.

In addition to Nemesis, we reimplemented and extended the original Rache~\cite{otawose_sigmod23} to support floating-point numbers. The key idea is to represent a floating-point number as its integral and decimal parts, both of which can be reconstructed using radixes. This enhanced version is referred to as Rache+ in our work.

\subsection{Experimental Setup}

In our experiments, we evaluate the proposed methods using the model weights trained from three widely adopted datasets: MNIST~\cite{lecun-mnist}, FashionMNIST~\cite{xiao2017fashion}, and CIFAR-10~\cite{krizhevsky2009learning}. The MNIST dataset consists of 70,000 grayscale images of handwritten digits (0--9), each with a resolution of $28 \times 28$ pixels. It is divided into 60,000 training samples and 10,000 testing samples.FashionMNIST is a dataset designed as a drop-in replacement for MNIST. It contains $28 \times 28$ grayscale images of clothing items, divided into 10 categories, such as T-shirts, trousers, and shoes. The dataset comprises 60,000 training samples and 10,000 testing samples. The CIFAR-10 dataset consists of 60,000 color images divided into 10 classes, including vehicles (e.g., airplanes, cars, trucks) and animals (e.g., cats, dogs, frogs). Each image has a resolution of $32 \times 32$ pixels and is represented using three color channels (RGB). The dataset is split into 50,000 training samples and 10,000 testing samples.

For MNIST and FashionMNIST, our network structure is a Convolutional Neural Network (CNN) with two convolutional layers and two fully connected layers. The first convolutional layer extracts low-level features with 32 channels and applies MaxPooling for downsampling. The second convolutional layer further refines these features with 64 channels and additional pooling. The fully connected layers map the extracted features to 512 hidden units and finally produce logits for a 10-class classification task. ReLU activations are applied throughout the network to introduce non-linearity and improve learning capacity. The overall number of weights and parameters is 582,026.

For CIFAR-10, our network structure is a CNN designed for image classification tasks on RGB datasets. The model consists of two convolutional layers and two fully connected layers. The convolutional layers extract hierarchical features from input images, with ReLU activation and MaxPooling applied for non-linearity and spatial downsampling. The fully connected layers map these features to a lower-dimensional space and output logits for a 10-class classification task. This structure is specifically adapted for RGB image datasets such as CIFAR-10, leveraging the increased input dimensionality to enhance feature extraction. The overall number of weights and parameters is 878,538.

\begin{table*}[t!]
\centering
\renewcommand{\arraystretch}{1.2}
\caption{Feature Comparison of State-of-the-art FHE Schemes}
\label{tab:comparison}
\begin{tabular}{lcccc}
\hline
 & Nemesis (this work) & Rache (2023) & CKKS (2017) & BGV/BFV (2012) \\ \hline
Supported Data Types & Float + Integer & Integer only & Float + Integer & Integer only \\ 
Support of Multiple Slots & $\checkmark$ & $\times$ & $\checkmark$ & $\checkmark$ \\
Precomputation/Encoding & \boldmath$O(1)$ & \boldmath$O(r)$ & \boldmath$O(d^k)$ & \boldmath$O(d^k)$ \\ 
Randomness Algorithm & Polynomial noise & Radix reshuffle & Additive zero  & Additive zero \\ 
Encryption Complexity & \boldmath$O(d)$ & \boldmath$O(rd)$ & \boldmath$O(d)$ & \boldmath$O(d)$ \\ 
Storage Overhead & \boldmath$O(1)$ & \boldmath$O(r)$ & \boldmath$O(1)$ & \boldmath$O(1)$ \\ 
\hline
\end{tabular}
\end{table*}

All of the experiments were conducted on the NSF Chameleon Cloud~\cite{keahey2020lessons} platform, utilizing a high-performance compute node. The machine is powered by two Intel Xeon E5-2670 v3 CPUs, operating at a base clock speed of 2.30 GHz. Together, the processors provide a total of 48 hardware threads. The CPU features a hierarchical cache structure, including 32 KB L1 data and instruction caches, 256 KB L2 caches per core, and a shared 30 MB L3 cache per processor. These specifications ensure efficient handling of computation-intensive tasks.
The system is equipped with 128 GiB of RAM, sufficient for processing large datasets and performing memory-intensive operations associated with compression and decompression experiments. Storage is provided by a 250 GB Seagate SATA SSD, which offers low-latency access to data files.
For GPU-based computations, the node includes two NVIDIA Tesla P100 GPUs, renowned for their high parallel processing capabilities.

In our experiments, we adopt the standard FedAvg~\cite{mcmahan2017fedavg} algorithm as the federated learning aggregation method. The global model is trained for 10 communication rounds (global rounds), with local model updates performed on each participating client during every round. All experiments are repeated three times, and the results are averaged to ensure robustness and mitigate potential randomness. Each client trains the model locally for 5 epochs per round using a batch size of 32, and the learning rate is set to 0.01 with no learning rate decay. A total of 20 clients participate in the training process, with a join ratio of 1.0, meaning all clients contribute to every communication round. The default CKKS ring degree is 4,096.

\subsection{Feature Comparison with State-of-the-arts}

Table~\ref{tab:comparison} compares Nemesis, Rache~\cite{otawose_sigmod23}, CKKS~\cite{ckks}, and BGV~\cite{bgv}/BFV~\cite{bfv} encryption schemes across multiple dimensions, including supported data types, multiple slot support, precomputation cost, randomness algorithms, randomness complexity, and storage overhead.

Nemesis and CKKS support both floating-point and integer data types, whereas Rache and BGV/BFV are restricted to integer data only. Unlike Rache, which does not support multiple slots, Nemesis, CKKS, and BGV/BFV provide this functionality. In terms of precomputation/encoding cost, Nemesis achieves $O(1)$ complexity, significantly outperforming CKKS's $O(d^k)$ and Rache's $O(r)$, where $d$ is the polynomial degree and $k$ reflects the homomorphic operation depth.

For randomness generation, Nemesis and CKKS employ a polynomial algorithm, while Rache relies on a more costly radix reshuffle approach. Nemesis achieves randomness complexity of $O(d)$, matching CKKS and BGV/BFV, while Rache incurs a higher complexity of $O(rd)$. In terms of storage overhead, Nemesis maintains minimal overhead at $O(1)$, similar to CKKS and BGV/BFV, whereas Rache introduces $O(r)$ overhead due to its radix-based design.

\subsection{Performance Comparison with State-of-the-arts}

\begin{figure}[t!]
    \centering
    \includegraphics[width=\columnwidth]{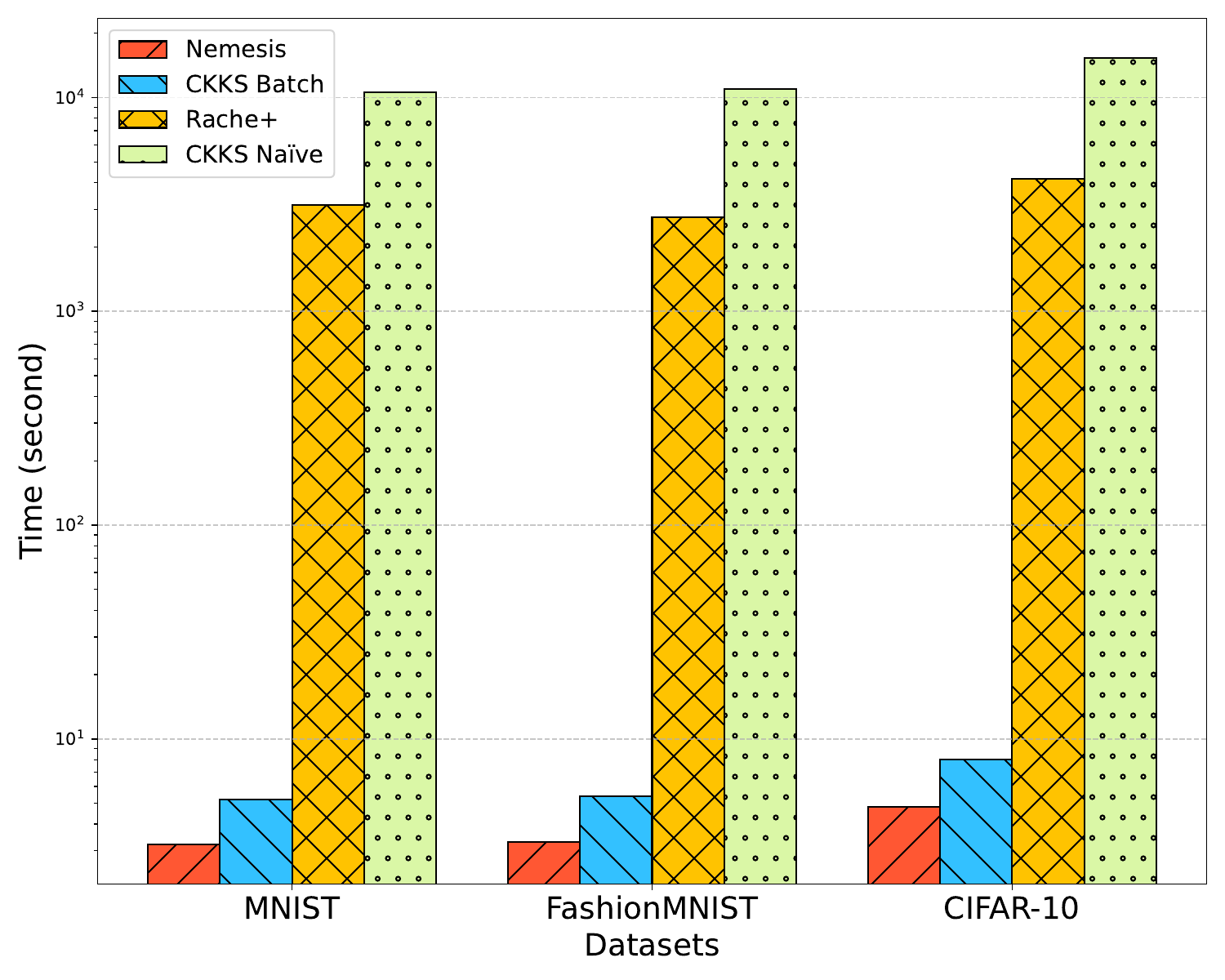}
    \caption{Comparison of Nemesis, CKKS Batch, Rache, and CKKS Naïve methods.}
    \label{fig:comparison_plot}
\end{figure}

Figure~\ref{fig:comparison_plot} compares the performance of four encryption schemes: Nemesis, CKKS Batch, Rache+, and CKKS Naive. The experiments were conducted using models trained on three datasets (MNIST, FashionMNIST, CIFAR-10) through federated learning with the FedAvg aggregation algorithm~\cite{mcmahan2017fedavg}. The y-axis shows the execution time in seconds, measured using a logarithmic scale to account for significant differences in execution time between the schemes.
CKKS Batch refers to the use of multiple slots for parallel encryption, allowing multiple data points to be packed and processed together efficiently. In contrast, CKKS Naive encrypts data points individually, leading to significantly higher computational overhead. The batch size is set to the maximum number of slots, which corresponds to half of the polynomial ring degree, i.e., $\frac{4096}{2} = 2048$. The goal of this figure is to provide a visual comparison of computation efficiency for each method across varying datasets.

Our results show that the Nemesis method achieves significantly lower computation times compared to CKKS Batch, with Nemesis being approximately 1.6 to 1.7 times faster across all datasets. Specifically, on CIFAR-10, Nemesis completes in 4.8 seconds, while CKKS Batch requires 8.0 seconds. Similarly, on MNIST and FashionMNIST, Nemesis achieves 3.2 and 3.3 seconds, compared to 5.2 and 5.4 seconds for CKKS Batch, respectively. In contrast, the Rache+ and CKKS Naive methods exhibit much higher execution costs, particularly on CIFAR-10, where Rache+ requires over 4,000 seconds and the Naive scheme exceeds 15,000 seconds. These results highlight the computational efficiency of Nemesis over the other methods, making it a superior choice for practical machine learning tasks.

\subsection{Nemesis: Precomputation}

The purpose of this experiment is to evaluate the time overhead introduced by precomputation during the encryption process. Specifically, we compare the execution time for plaintext packing and batch encryption across different batch sizes. The batch size ranges from 1 to 2,048 (recall the polynomial ring degree is 4,096), which represents the full range of slots utilized in the CKKS scheme. The experiment helps identify how precomputation scales with increasing batch sizes and provides insight into its overall efficiency in practice.

\begin{figure}[t!]
    \centering
    \includegraphics[width=\linewidth]{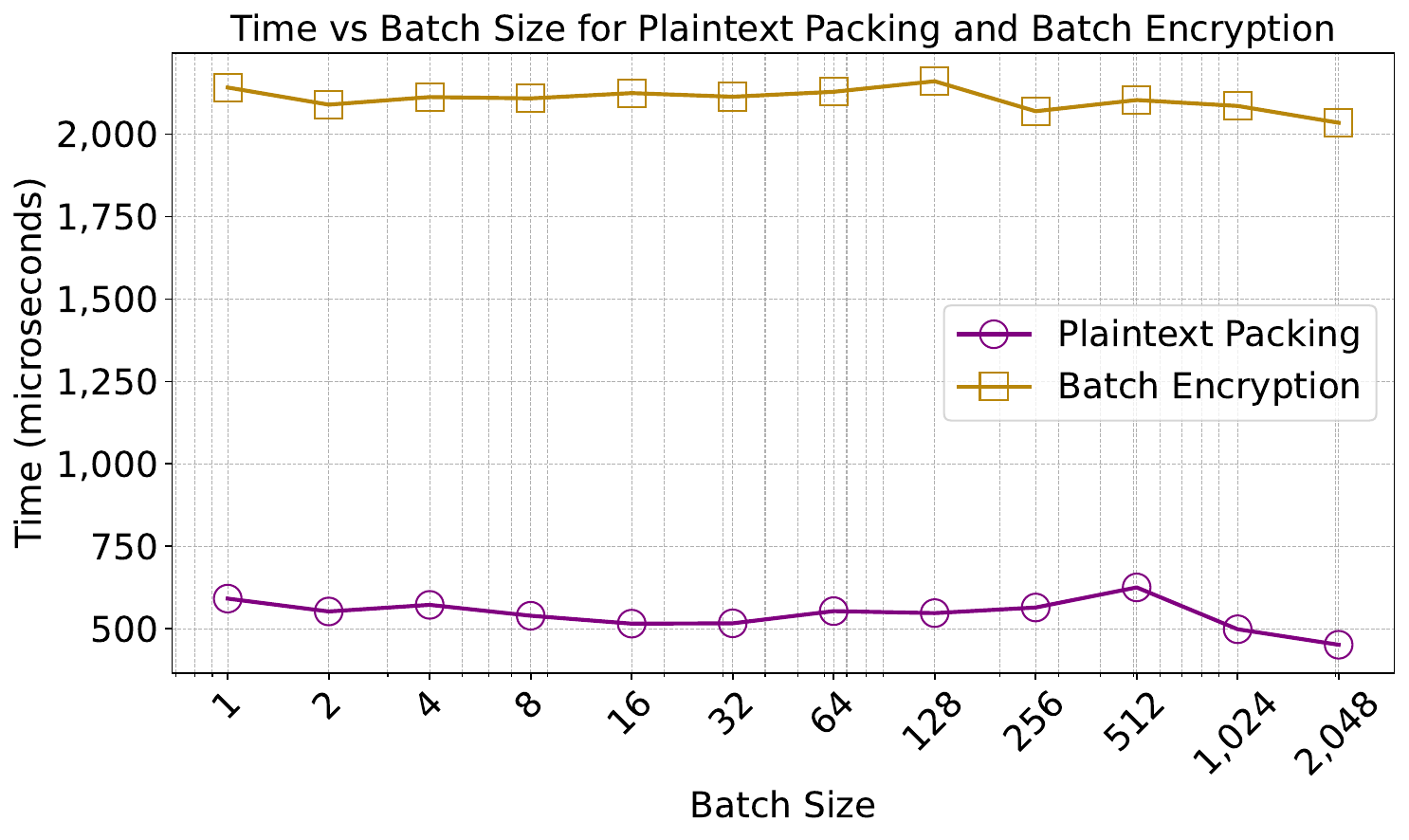}
    \caption{Nemesis' Time Overhead (i.e., precomputation) for Plaintext Packing and Batch Encryption across Different Batch Sizes}
    \label{fig:nemesis_overhead}
\end{figure}

From the results shown in Figure~\ref{fig:nemesis_overhead}, we observe that the time for plaintext packing and batch encryption remains relatively stable across all batch sizes, with minor fluctuations. This suggests that the precomputing overhead of applying Nemesis is not significantly affected by the batch size. The reason for this stability lies in the nature of the encoding and encryption process. While the batch size determines the number of plaintexts combined into a single polynomial, the transformation required to encode different numbers of plaintexts into one polynomial incurs only negligible performance overhead. Furthermore, the encryption of the resulting polynomial—irrespective of the number of plaintexts it represents—relies primarily on the fixed computational complexity of the encryption scheme itself, which remains largely unaffected by changes in batch size. Consequently, the overall time required for plaintext packing and encryption exhibits minimal dependency on the batch size. These insights are critical for optimizing the performance of systems leveraging the CKKS scheme in practical scenarios.

\subsection{Nemesis: Reconstruction}

To evaluate the performance overhead of ciphertext-batch reconstruction, we conducted experiments with the CNN models trained from three datasets. For each dataset, we measured the reconstruction time when processing varying batch sizes, specifically \{128, 256, 512, 1024, 2048\}. The goal of this experiment is to quantify how reconstruction time scales with batch size and to compare the relative performance across different datasets. Reconstruction involves chunking plaintext batches, performing homomorphic computations across the plaintext-space and the ciphertext-space, and managing discrete Gaussian noises efficiently.

\begin{figure}[t!]
    \centering
    \includegraphics[width=\linewidth]{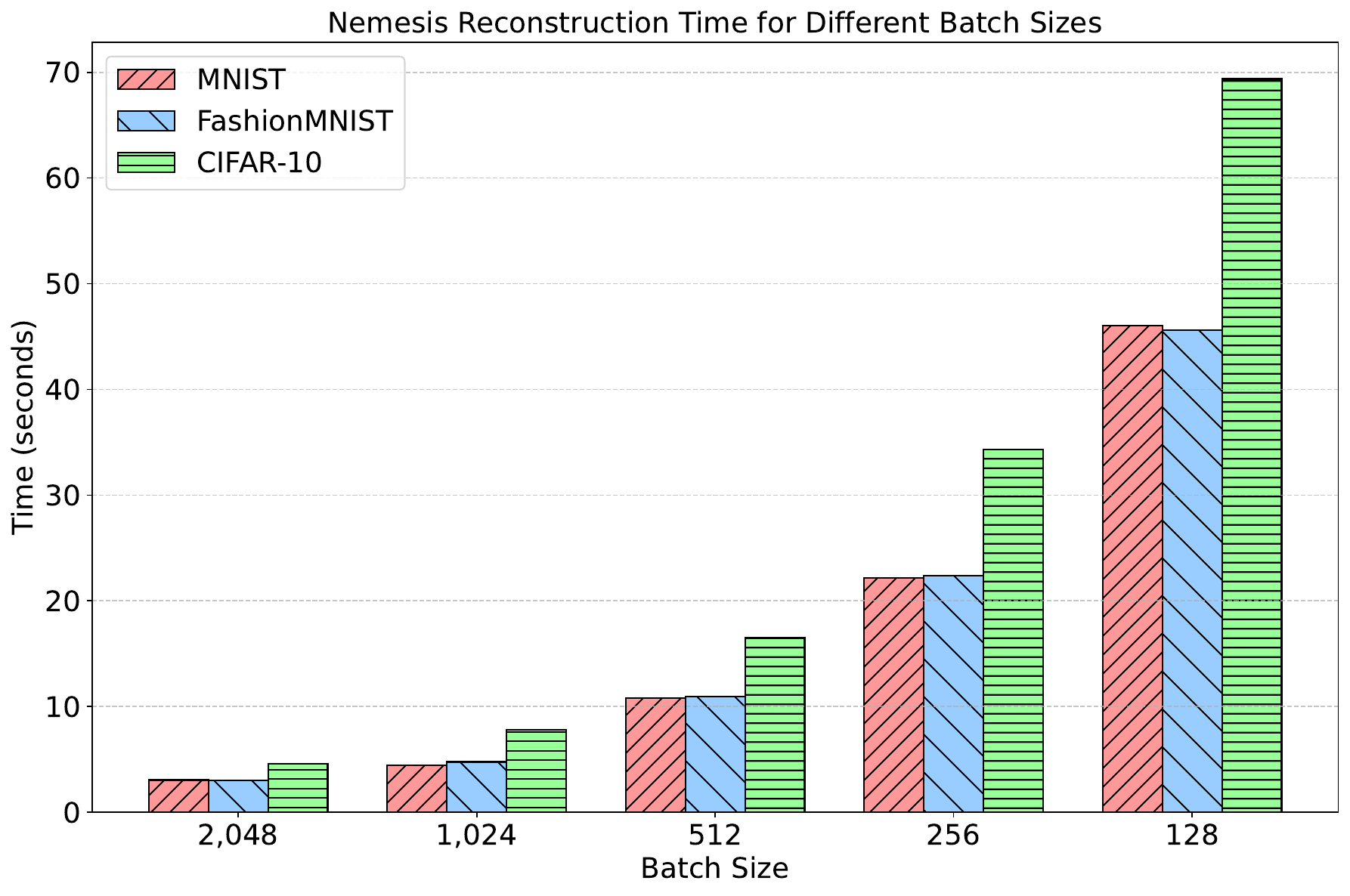}
    \caption{Nemesis' Reconstruction Time with Different Batch Sizes.}
    \label{fig:reconstruct}
\end{figure}

Figure~\ref{fig:reconstruct} shows that the reconstruction time increases as the batch size decreases, which is consistent across all three datasets. This trend indicates that smaller batch sizes incur higher relative overhead due to larger number of batches. Among the datasets, CIFAR-10 consistently shows the highest reconstruction time. This difference is attributed to the larger complexity and higher computational cost of CIFAR-10's CNN model compared to the others.

\subsection{Nemesis: Randomization}

To evaluate the computational overhead introduced by polynomial-noise randomization, we conducted experiments with batch sizes in \{128, 256, 512, 1024, 2048\}. In this experiment, we measured the time required to generate and apply random noise polynomials during ciphertext processing. This process is critical for maintaining security in homomorphic encryption while ensuring computational efficiency.

\begin{figure}[t!]
    \centering
    \includegraphics[width=\linewidth]{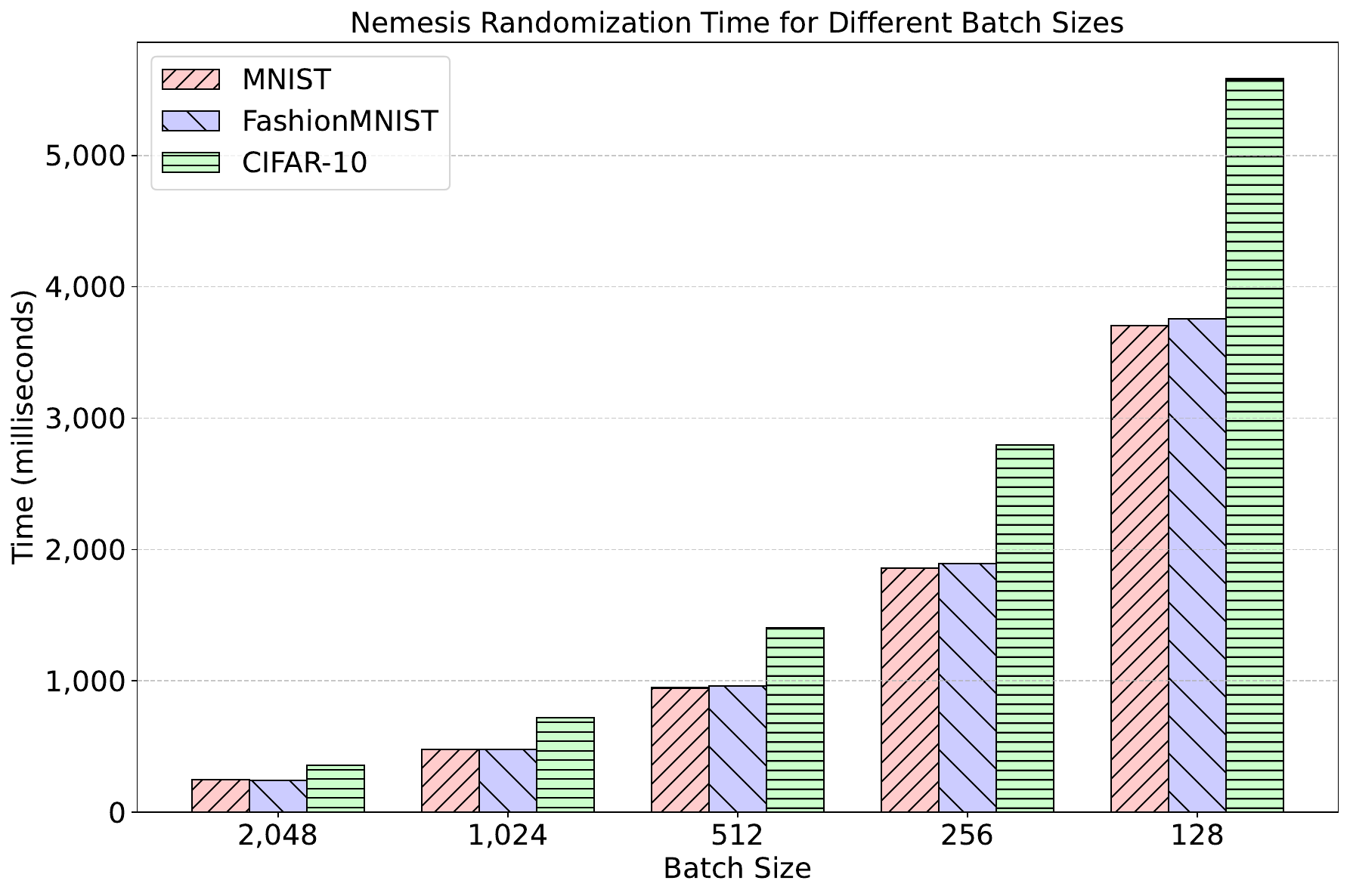}
    \caption{Nemesis' Randomization Time with Different Batch Sizes.}
    \label{fig:randomization}
\end{figure}

Figure~\ref{fig:randomization} shows that the time required for polynomial-noise randomization increases significantly as the batch size decreases. This trend is consistent across all three datasets, highlighting the impact of batch size on computational overhead. Among the datasets, CIFAR-10 exhibits the highest randomization time, reflecting its greater complexity and larger data representation compared to MNIST and FashionMNIST in terms of their trained CNN models. 

\subsection{Nemesis: Time Distribution}

In this experiment, we evaluate the end-to-end time distribution of the three core stages of the Nemesis system: Precomputation, Reconstruction, and Randomization. The goal is to determine the relative overhead of each stage across varying batch sizes, specifically focusing on batch sizes ranging from 128 to 2048. Precomputation involves setting up initial ciphertexts, Reconstruction focuses on ciphertext-batch creation, and Randomization ensures secure polynomial noise injection into ciphertexts. The percentage breakdown of these stages is illustrated in the stacked histogram, where each batch size has been normalized to 100\% for easier comparison.

\begin{figure}[t!]
    \centering
    \includegraphics[width=\linewidth]{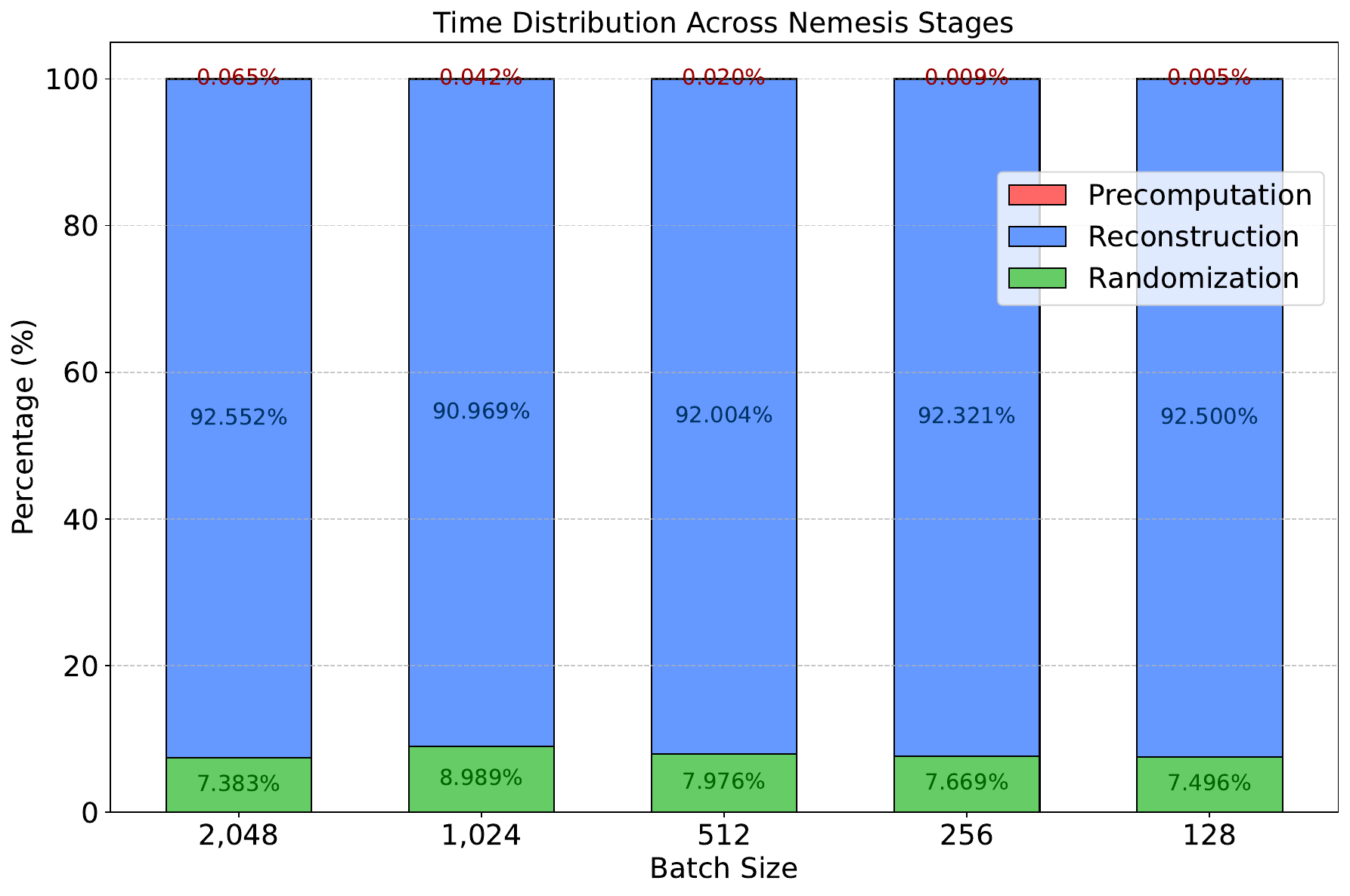}
    \caption{End-to-end time distribution across Nemesis stages for varying batch sizes.}
    \label{fig:end_to_end_distribution}
\end{figure}

As seen in Figure~\ref{fig:end_to_end_distribution}, Precomputation overhead is negligible (less than 0.1\%) across all batch sizes, confirming its minimal contribution to the overall time. Reconstruction dominates the total time, accounting for over 90\% in all cases, regardless of batch size. Randomization contributes a small but noticeable portion, ranging between 7.4\% and 9.0\%. These results highlight that optimization efforts, possibly in the future work alone this line of research, should primarily focus on the Reconstruction stage to achieve significant end-to-end performance improvements, while Randomization remains consistent and manageable.

\section{Conclusion}

In this work, we tackled the computational inefficiency of Fully Homomorphic Encryption (FHE), a major bottleneck in privacy-preserving machine learning systems. We introduced \textit{Nemesis}, a caching-based framework that enhances FHE performance without compromising security or accuracy. Leveraging advanced mathematical tools, Nemesis extends prior works, like Rache, by supporting multi-slot encryption and general plaintext structures. Formal analysis proves that Nemesis maintains semantic security under standard cryptographic assumptions. Extensive experiments on datasets like MNIST, FashionMNIST, and CIFAR-10, using \texttt{OpenFHE} and \texttt{PFLlib}, show significant performance gains, establishing Nemesis as a practical FHE solution.

Future work may extend \textit{Nemesis} to broader cryptographic settings, such as integrating with advanced machine learning models for improved scalability in privacy-preserving tasks. Supporting dynamic ciphertext updates could enable real-time encrypted computations, while exploring hybrid approaches like combining FHE with secure multi-party computation may unlock new possibilities for collaborative analytics.

\section*{Impact Statement}

This paper presents work whose goal is to advance the field of 
Machine Learning. There are many potential societal consequences 
of our work, none which we feel must be specifically highlighted here.

\section*{Acknowledgment}
Results presented in this paper were partly obtained using the Chameleon testbed supported by the National Science Foundation. We also appreciate the insightful discussions and feedback from Professor Stefano Tessaro at the University of Washington, which helped refine some of the key ideas presented in this paper.


\bibliography{example_paper}
\bibliographystyle{icml2025}




\end{document}